# ANALYZING CRYPTOGRAPHIC ALGORITHMS FOR SECURE CLOUD NETWORK

Vineet Kumar Singh[1]
M.E. Student[1]
[1]Department of Computer Science NITTTR
Chandigarh (U.T.), India

Dr. Maitreyee Dutta[2]
Associate Professor[2]
[2]Department of Computer Science, NITTTR
Chandigarh (U.T.), India

*Abstract*—**Pay as per usage concept of Cloud computing has brought revolutionary changes in the information technology world. Cloud computing is now the need of changing technological scenario and even more service consumers are adopting it every day. One of the most fascinating features of cloud is that, from small scale to large scale, it is beneficial for all types of organizations. It has offered several advantages to the society, but at the same time many organizations are still reluctant towards the adoption of this technological innovation, and unfortunately the reason is true. The security concerns of cloud network are even increasing along with today's growing cloud service consumers. These security concerns can be resolved by using proper security mechanism. In this paper we have analyzed the performance of some popular cryptographic algorithms for the cloud network, namely; Symmetric Algorithms, Asymmetric Algorithms, Hash algorithms and Homomorphic Algorithms. Homomorphic Algorithms are relatively newer than other categories, but these algorithms will be having wide application in near future, especially in the untrusted environment like cloud computing. We have conducted the comparative study in between two most popular algorithms of each category; AES vs. DES, RSA vs. ELGAMAL, MD5 vs. SHA, Paillier vs. Benaloh, so that performance wise better algorithms can be concluded for the cloud network. The algorithms are tested on the single system and on the cloud environment as well.**

## I. INTRODUCTION

Cloud computing is a new way of delivering the services based on pay as per usage model of computing, hence cloud computing is a new different way of delivering the services instead of completely new technology. In cloud computing, consumers can use huge amount of resources from cloud network, as per their need, at the same time malicious users may also use these huge computational resources of cloud and can launch an attack on the legitimate users. Thus, security is a major consideration in cloud as the owner does not control the data while the control lies in service providers hand [1].According to the Cloud Computing Services Survey, done by IDC IT group in 2009, over 87% of the people said that security is the number one issue which prevents the adoption of the cloud computing [2].

As cloud computing is a new enhanced version of the technology, it has also created some new challenges, which are quite different from traditional security challenges. Based on the survey of Cloud Security Alliance (CSA),"The Notorious Nine: Cloud Computing Top Threats in 2013"[3], which they performed with the industry experts on greatest vulnerabilities of cloud computing, there are nine critical threats to cloud security and among them "Data Breaches" is ranked as top threat .Data breaches are ranked as number one threat for the cloud computing. Since the inception of cloud computing technology, this threat is still present in the system. Multi-tenancy is one of the most important reasons among several, for data breaches. Since data from various organizations lie together in a multi-tenant cloud environment, breaching into the cloud will ultimately attack the data of all the users. Thus, the cloud of such huge information becomes an attractive target for attackers [4].

Data remanence is also one of the reasons of security breach, & generally it is unintentional. Data remanence is the vestigial of data that have been nominally removed or migrated. As several virtual machines running on one physical machine lack of separation between multiple users, may lead to the unwilling disclosure of private data in case of data remanence. This presents higher risk to the cloud customers than with dedicated hardware [5, 6].

Trusted third Party services within the cloud, establishes the necessary trust level and provides ideal solutions to preserve the confidentiality, integrity and authenticity of data and Communications [5, 7].Breach notifications are also important as Poor breach notification may lead to privacy breach [8]. Unfortunately, the breach notifications could not really protect a customer's data. A recent survey shows that service consumers who have received data breach notifications within the past year are at a much greater risk for fraud than the typical service consumers [9]. Daniel J. Abadi concluded [10] that it is a great risk in storing transactional data on an un-trusted host. Transactional databases contain the complete set of operational data needed to power overall business processes. This data includes detail at the lowest granularity, and often includes important information such as credit card numbers of the customers. Thus, any increase in potential security breaches is typically unacceptable. Facebook user data breach is a recent example of the questionable user data safety on cloud systems [11].Cloud is a very huge information repository and no client would like to take risk on his/her information. If we consider that cloud end security is up to the mark, then also there is question about unsecure client data while transition to the cloud end. Our next section is an attempt to find out some efficient solutions, which can be helpful in real time environment.

## II. Proposed Evaluation Algorithms

Cloud computing involves frequent uploading and downloading of data along with heavy computation on servers which are managed by third party. Since the client does not control the cloud environment, always there is a probability of losing the confidentiality and integrity of data either by intentional or unintentional means. It is very important to use required cryptographic algorithms to save client data on cloud network for. Thus, if cloud providers & users are aware in the selection of proper cryptographic algorithms as per their performance in different executing environments on different types of input sizes & different security needs, then only high cloud security without performance degradation can be achieved, and in turn cloud adoption will be increased. Thus for the sake of obtaining the high performance cryptographic algorithms for better cloud security, an analytical performance analysis between different cryptographic algorithms is done on the basis of Total Execution Time of the algorithms and their Speed-Up Ratios on varying Input Size. Performance analysis is done for the most popularly used cryptography algorithms in the following fashion-

- ➢ AES vs. DES Symmetric Algorithms
- ➢ RSA vs. ELGAMAL Asymmetric Algorithms
- ➢ MD5 vs. SHA Encoding Algorithms
- ➢ BENALOH vs. PAILLIER Partial Homomorphic Algorithms

It is of utmost importance to understand the computational overhead for various algorithms and recommend appropriate schemes under different situation. In the present research work, some widely used symmetric, asymmetric, encoding and homomorphic cryptographic techniques are analyzed and compared on the basis of some factors viz. file size, total execution time, speed up ratio. The proposed algorithms for performance analysis are following:-

AES (Advance Encryption Standard):

AES is a symmetric block cipher. This means that it uses the same key for both encryption and decryption. However, AES is quite different from DES in a number of ways. The algorithm Rijndael allows for a variety of block and key sizes and not just the 64 and 56 bits of DES' block and key size. The block and key can in fact be chosen independently from 128, 160, 192, 224, 256 bits and need not be the same. At present the most common key size likely to be used is the 128 bit key, thus 128 bit key size is used for analysis.

DES (Data Encryption Standard):

The DES (Data Encryption Standard) is a block cipher. It encrypts data in blocks of size 64 bits each. That is 64 bits of plain text goes as input to DES, which produces 64 bits of cipher text. The same algorithm and key are used for encryption and decryption, with minor differences. The key length of this algorithm is 56 bits; however a 64 bits key is actually input. DES is therefore a symmetric key algorithm.

RSA (Rivest-Shamir-Adleman):

RSA (Rivest-Shamir-Adleman) is an algorithm for public-key cryptography, involves a public key and a private key. The public key can be known to everyone and is used for encrypting messages. Messages encrypted with the public key can only be decrypted using the private key. It protected user data include encryption prior to storage, user authentication procedures prior to storage or retrieval, and building secure channels for data transmission.4096 bit key size is most popular as it is used for encrypting the symmetric keys, thus 4096 bit key size is used for execution of RSA algorithm.

The ElGamal Cryptosystem:

This algorithm belongs to the family of public key cryptographic algorithms. Therefore it makes use of a key separated into a public and a private part. A fundamental aspect of this system is that the knowledge of the private part makes the decryption easy. 4096 bit key size is used for performance analysis.

MD5 (Message Digest5):

MD5 (Message Digest5) is a widely used cryptographic hash function with a 128-bit hash value It processes a variable-size message into a fixed-length output of 128 bits. The input message is divided into chunks of 512-bit blocks; then the message is padded for making its length divisible by 512. In this sender use the public key of the receiver to encrypt the message and receiver use its private key to decrypt the message.

SHA (Secure Hashing Algorithm):

SHA stands for "Secure Hashing Algorithm". It is a hashing algorithm designed by the United States National Security Agency and published by NIST. It is the improvement upon original SHA and was first published in 1995. SHA-1 is most widely used SHA hash function, but very soon it is going to be replaced by the newer and stronger SHA-2 hash function. It is currently used in a wide variety of applications, including TLS, SSL, SSH and PGP.SHA1 outputs a 160-bit digest of any sized file or input.

Benaloh Homomorphic Cryptosystem:

The Benaloh Cryptosystem is an extended version of the GM cryptosystem created in 1994 by Josh Benaloh. The main advantage of the Benaloh Cryptosystem over GM is that longer blocks of input message can be encrypted at once, whereas in GM each bit is encrypted separately, and the encryption cost is also not too high.

Paillier's Homomorphic Cryptosystem:

The Paillier scheme was first published by Pascal Paillier in 1999. Paillier probabilistic scheme has created a good amount of interest and further study since it was originated. The main interest seems to be centered around another property it possesses: the homomorphic property allows this scheme to do normal addition operations on several encrypted values and achieving the encrypted sum, The encrypted sum can be decrypted later without even knowing the values ever, that made up the sum.

III. Parameters Used for Problem Analysis

**Total Execution Time**

Total Execution time is the sum of the total time taken during the homomorphic operation including the key generation time, encryption time and decryption time.

**Speed-Up Ratio**

Speed-Up Ratio is defined as the ratio of Total Execution Time on a local processor to the Total Execution Time on the cloud network. It will also be evaluated both on local system and on cloud network.

**Input Size:**

The performance analysis of all the algorithms will be done on the basis of different input file sizes on the local system and on cloud network as well. Algorithms performance varies heavily on different input sizes of the plain text.

IV. Proposed Methodology & Execution Environment

In the proposed methodology, performance analysis of the given algorithms on the basis of the parameters- Total Execution Time and Speed-Up Ratio and is done on local system as well as on the cloud network on varying Input Size.JavaSE-1.7 on Eclipse SDK 4.3.0 release is used for the development of all the algorithms.

Cloud software environment provider supplies the developers with programming-level-environment with well defined set of API's. Cloud-enabled applications on Spoon allow software developers to make their existing desktop applications available in the cloud, with no installs. Spoon applications can be accessed from the Spoon.net library. Spoon offers many software through their SAAS offerings, we used Eclipse 4.3.0 cloud SAAS for executing all the java algorithms in cloud environment. All the algorithms are tested on Intel core i5 third generation processor with MS Windows 7 Home Premium 64 bit SP-1.Processor speed is 2.50 GHz and with 2 GB RAM.

## V. Observation Results

All the results are obtained with due care , for achieving higher accuracy five samples of Total Execution Time were taken then an average of five samples were taken for the measurement and comparative analysis among algorithms and for the graph plotting as well. All the respective observation readings and graph are shown for all the analyzed algorithms on single system and on cloud network as well.

1) Total Execution Time for Single System

| ALGO / INPUT SIZE | SAMPLE-1 | SAMPLE-2 | SAMPLE-3 | SAMPLE-4 | SAMPLE-5 | AVERAGE |
|---|---|---|---|---|---|---|
| AES(10 KB) | 234 | 265 | 312 | 249 | 250 | **262** |
| AES(20 KB) | 277 | 284 | 252 | 245 | 262 | **264** |
| AES(30 KB) | 260 | 266 | 273 | 263 | 275 | **267** |
| AES(40 KB) | 288 | 271 | 272 | 273 | 271 | **275** |
| AES(50 KB) | 276 | 308 | 276 | 306 | 291 | **291** |
| AES Average Total Execution Time(ms) | | | | | | **272** |
| DES(10 KB) | 239 | 261 | 258 | 246 | 247 | **250** |
| DES(20 KB) | 252 | 250 | 260 | 254 | 256 | **254** |
| DES(30 KB) | 246 | 430 | 251 | 266 | 259 | **290** |
| DES(40 KB) | 267 | 251 | 280 | 327 | 337 | **292** |
| DES(50 KB) | 297 | 278 | 261 | 319 | 331 | **297** |
| DES Average Total Execution Time(ms) | | | | | | **277** |
| ELGAMAL (100 B) | 497 | 369 | 489 | 463 | 397 | **443** |
| ELGAMAL (200 B) | 391 | 471 | 463 | 340 | 548 | **443** |
| ELGAMAL (300 B) | 457 | 458 | 419 | 468 | 440 | **448** |
| ELGAMAL (400 B) | 382 | 406 | 439 | 417 | 608 | **450** |
| ELGAMAL (501 B) | 459 | 463 | 590 | 531 | 534 | **515** |
| Average Total Execution Time(ms) | | | | | | **460** |
| RSA(100 B) | 2376 | 2516 | 2902 | 3547 | 3022 | **2873** |
| RSA(200 B) | 4725 | 4516 | 3722 | 4014 | 1786 | **3753** |
| RSA(300 B) | 5431 | 5423 | 3427 | 4616 | 4922 | **4764** |
| RSA(400 B) | 6982 | 6764 | 4926 | 5531 | 3110 | **5463** |
| RSA(501 B) | 6725 | 8095 | 6551 | 5986 | 4066 | **6284** |
| Average Total Execution Time(ms) | | | | | | **4627** |
| MD5(10 KB) | 14 | 19 | 19 | 18 | 16 | **17** |
| MD5(20 KB) | 19 | 20 | 17 | 13 | 23 | **18** |
| MD5(30 KB) | 21 | 20 | 21 | 21 | 21 | **21** |
| MD5(40 KB) | 22 | 23 | 22 | 26 | 17 | **22** |
| MD5(50 KB) | 23 | 23 | 21 | 17 | 24 | **22** |
| MD5 Average Total Execution Time(ms) | | | | | | **20** |
| SHA(10 KB) | 18 | 20 | 20 | 18 | 20 | **19** |
| SHA(20 KB) | 28 | 27 | 29 | 30 | 31 | **29** |
| SHA(30 KB) | 31 | 38 | 38 | 32 | 30 | **34** |
| SHA(40 KB) | 32 | 31 | 38 | 37 | 40 | **36** |
| SHA(50 KB) | 37 | 38 | 37 | 39 | 37 | **38** |
| SHA Average Total Execution Time(ms) | | | | | | **31** |

| Execution Speed of Homomorphic Algorithms | | | | | | |
|---|---|---|---|---|---|---|
| **PAILLIER** | 344 | 330 | 360 | 356 | 337 | **345** |
| Paillier Average Total Execution Time(ms) | | | | | | **345** |
| **BENALOH** | 667 | 556 | 600 | 625 | 583 | **606** |
| Benaloh Average Total Execution Time(ms) | | | | | | **606** |

2) Total Execution Time for Cloud Network

| ALGO / INPUT SIZE | SAMPLE- 1 | SAMPLE -2 | SAMPLE- 3 | SAMPLE -4 | SAMPLE- 5 | AVERAGE |
|---|---|---|---|---|---|---|
| AES(10 KB) | 239 | 234 | 228 | 236 | 238 | **235** |
| AES(20 KB) | 238 | 232 | 237 | 233 | 243 | **237** |
| AES(30 KB) | 240 | 243 | 240 | 235 | 236 | **239** |
| AES(40 KB) | 249 | 260 | 245 | 242 | 241 | **247** |
| AES(50 KB) | 258 | 250 | 243 | 247 | 254 | **250** |
| AES Average Total Execution Time(ms) | | | | | | **242** |
| DES(10 KB) | 240 | 236 | 233 | 240 | 239 | **238** |
| DES(20 KB) | 239 | 236 | 239 | 235 | 242 | **238** |
| DES(30 KB) | 238 | 234 | 244 | 247 | 236 | **240** |
| DES(40 KB) | 247 | 246 | 255 | 252 | 257 | **251** |
| DES(50 KB) | 249 | 268 | 254 | 267 | 264 | **260** |
| DES Average Total Execution Time(ms) | | | | | | **245** |
| ELGAMAL (100 B) | 301 | 187 | 236 | 185 | 362 | **254** |
| ELGAMAL (200 B) | 289 | 283 | 465 | 708 | 403 | **430** |
| ELGAMAL (300 B) | 542 | 221 | 566 | 226 | 666 | **444** |
| ELGAMAL (400 B) | 509 | 390 | 414 | 465 | 461 | **447** |
| ELGAMAL (501 B) | 500 | 600 | 493 | 423 | 466 | **496** |
| Average Total Execution Time(ms) | | | | | | **414** |
| RSA(100 B) | 5469 | 4916 | 5189 | 4356 | 5643 | **5115** |
| RSA(200 B) | 7985 | 8734 | 7835 | 6534 | 7856 | **7789** |
| RSA(300 B) | 8453 | 7649 | 8634 | 8964 | 8857 | **8511** |
| RSA(400 B) | 9089 | 9954 | 10067 | 10075 | 10674 | **9972** |
| RSA(501 B) | 11589 | 12563 | 12775 | 12885 | 13437 | **12650** |
| Average Total Execution Time(ms) | | | | | | **8807** |
| MD5(10 KB) | 17 | 16 | 17 | 17 | 17 | **17** |
| MD5(20 KB) | 18 | 18 | 18 | 19 | 19 | **18** |
| MD5(30 KB) | 18 | 19 | 21 | 18 | 19 | **19** |
| MD5(40 KB) | 19 | 20 | 19 | 20 | 18 | **19** |
| MD5(50 KB) | 21 | 21 | 20 | 21 | 21 | **21** |
| MD5 Average Total Execution Time(ms) | | | | | | **19** |
| SHA(10 KB) | 19 | 19 | 20 | 19 | 19 | **19** |
| SHA(20 KB) | 19 | 20 | 21 | 21 | 22 | **21** |
| SHA(30 KB) | 20 | 22 | 23 | 22 | 23 | **22** |
| SHA(40 KB) | 20 | 22 | 23 | 23 | 22 | **22** |
| SHA(50 KB) | 22 | 22 | 22 | 23 | 23 | **22** |
| SHA Average Total Execution Time(ms) | | | | | | **21** |

| Execution Speed of Homomorphic Algorithms | | | | | | |
|---|---|---|---|---|---|---|
| **PAILLIER** | 708 | 668 | 925 | 718 | 681 | **740** |
| Paillier Average Total Execution Time(ms) | | | | | | 740 |
| **BENALOH** | 1145 | 1253 | 1137 | 1461 | 1225 | **1244** |
| Benaloh Average Total Execution Time(ms) | | | | | | 1244 |

3) Speed Up Ratio for Different Algorithms

| AES | DES | ELGAMAL | RSA | MD5 | SHA | PAILLIER | BENALOH |
|---|---|---|---|---|---|---|---|
| 1.12 | 1.13 | 1.11 | 0.52 | 1.05 | 1.47 | 0.46 | 0.48 |

## VI. Outcomes of the Research

### AES vs. DES Symmetric Algorithms

- Both algorithms are found considerably fast on cloud network as compared to local system.
- Both algorithms achieved almost equal speed -Up Ratio, greater than one, which indicates fast processing over cloud network.
- AES is found to be faster than DES on both the local system and on the cloud network.

### RSA vs. ELGAMAL Asymmetric Algorithms

- Elgamal is found considerably fast over cloud network as compared to local system.
- RSA is found to be very slow on local system as well as on the cloud network.
- Elgamal achieved greater than one Speed-Up Ratio, which verifies its fast operation on cloud.RSA needs extra processing power for its fast operation from cloud network, it needs better configuration machines (more number of processors, fast processors, more RAM and cache memory) to operate than other cryptographic algorithms, as its speed Up Ratio is less than one.
- Elgamal is considerably fast than RSA on the local system and on the cloud network as well.

### MD5 vs. SHA Encoding Algorithms

- MD5 is found little bit fast on cloud network as compared to local system.
- SHA is found considerably fast on cloud network as compared to local system.
- MD5 achieves just greater than one Speed-Up Ratio which indicates higher speed on cloud network.
- SHA achieved a good speed-Up Ratio, as compared to MD5, which verifies its fast operation on cloud network.

### Benaloh vs. Paillier Homomorphic Algorithms

- Paillier is considerably fast than Benaloh on local system and on cloud network as well.
- Both algorithms are found very slow on cloud network as compared to local system.
- Both algorithms achieved almost equal speed –Up Ratio, less than one, almost half, which indicates their slow processing over cloud network.
- Both algorithm needs an extra processing power for its fast operation from cloud network, it needs better configuration machines (more number of processors, fast processors, more RAM and cache memory) to operate than other cryptographic algorithms.

## VII. Graphs of Different Algorithms

**Fig 1.** AES: Input Size vs. Total Execution Time

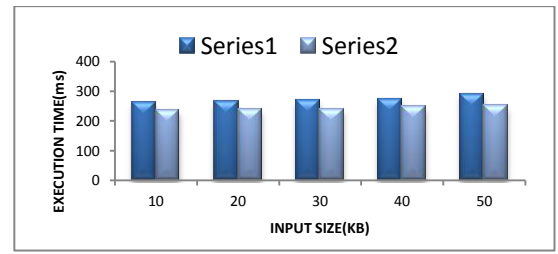

**Fig 2.** DES: Input Size vs. Total Execution Time

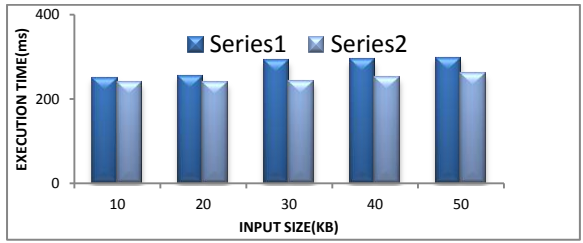

**Fig 3.** Elgamal: Input Size vs. Total Execution Time

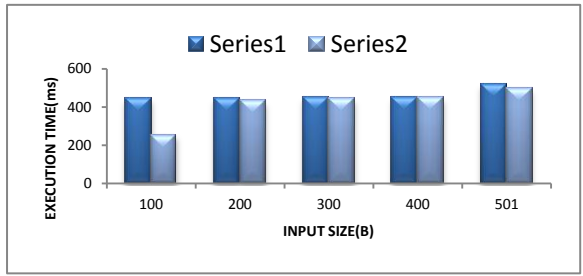

**Fig 4.** RSA: Input Size vs. Total Execution Time

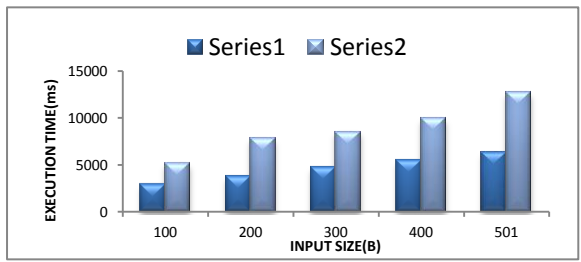

**Fig 5.** MD5: Input Size vs. Total Execution Time

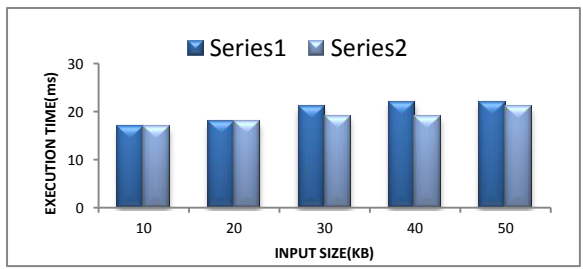

**Fig 6.** SHA: Input Size vs. Total Execution Time

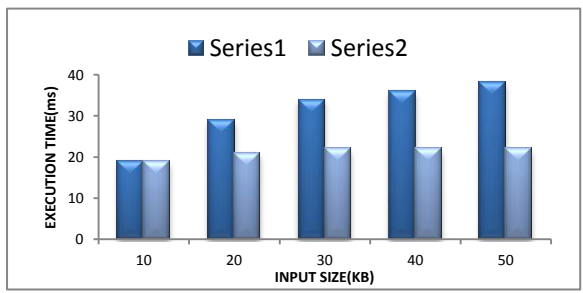

**Fig 7**. PAILLIER: Single System vs. Cloud Network

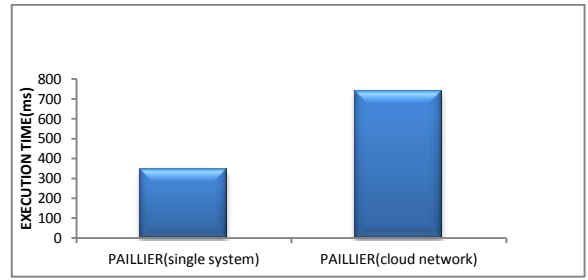

**Fig 8**. BENALOH: Single System vs. Cloud Network

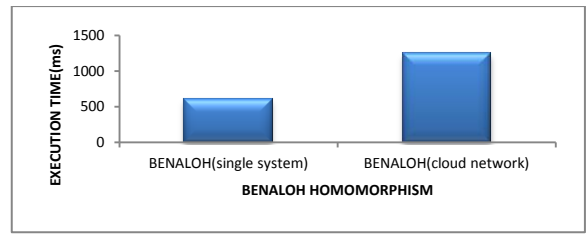

Let us understand what these graphs depict actually about the algorithms and their execution. In graph series 1 indicates single system while series 2 indicates cloud network.

Fig 1 illustrates the performance of AES algorithm on varying input size text files from 10 KB to 50 KB on single system as well as on cloud network.AES Total Execution Time increases gradually on single system while increasing input file size up-to file size of 40 KB. After that while the input is increased to 50 KB file size, the Total Execution Time also increased to large extent, from 275 ms to 291 ms. Cloud execution of AES achieves almost consistent increment in Total execution Time on different input size. Graph also depicts that in AES, Total Execution Time on Cloud network is significantly less than the Total Execution Time on single system on all the varying input size tested.

Fig 2 illustrates the performance of DES algorithm on varying input size text files from 10 KB to 50 KB on single system as well as on cloud network.DES Total Execution Time increases slowly initially from 10 KB to 20 KB, but increases rapidly on 30 KB input file size, and after then again increases slowly up-to 50 KB on single system.

Fig 3 illustrates the performance of Elgamal algorithm on varying input size text files from 100 B to 501 B on single system as well as on cloud network.DES Total Execution Time increases slowly initially from 100 KB to 400 KB, but increases sharply on 501 B file size. Cloud execution of Elgamal achieves almost consistent Total Execution

Time from 200 B to 400 B file size but increases rapidly on changing the file size from 100 B to 200 B and from 400 B to 500 B.

Fig 4 illustrates the performance of RSA algorithm on varying input size text files from 100 B to 501 B on single system as well as on cloud network. The graph of RSA is very interesting in the fact that RSA Total Execution Time is significantly higher than of Elgamal algorithm on both the single system as well as on cloud and on all varying input file size from 100 B to 500 B.RSA Total Execution Time increases rapidly on all the input size from 100 B to 500 B file size.

Fig 5 illustrates the performance of MD5 algorithm on varying input size text files from 10 KB to 50 KB on single system as well as on cloud network.MD5 Total Execution Times increases gradually on single system on increasing input file size up-to file size of 50 KB. Cloud execution of MD5 also achieves almost consistent slow increment in Total execution Time on different input size.

Fig 6 illustrates the performance of SHA algorithm on varying input size text files from 10 KB to 50 KB on single system as well as on cloud network.SHA Total Execution Times increases rapidly on single system on increasing input file size up-to file size from 10 KB to 20 KB. Then it increases gradually on increasing input size.

Fig 7 illustrates the performance of Paillier Homomorphic System on single system as well as on cloud network. Graph clearly indicates that Paillier homomorphic operation on single system is significantly fast from cloud network, and clearly indicates the need of extra processing power to operate homomorphic operations from cloud network.

Fig 8 illustrates the performance of Benaloh Homomorphic System on single system as well as on cloud network. Graph clearly indicates that Benaloh homomorphic operation on single system is significantly fast from cloud network, and clearly indicates the need of extra processing power to operate homomorphic operations from cloud network.

## VII. Conclusion & Future Scope

Earlier algorithms are implemented on local processor system, but now encryption and decryption techniques are implemented on cloud network too. It clearly indicates the need of more resilient algorithms for cloud network. Homomorphic cryptosystems have also added new challenges towards the secure and fast execution of programs on cloud. The results are obtained on the basis of Speed-Up Ratio and Total Execution Time parameter on varying input sizes. All the algorithms are applied on both the cloud network and local system. Comparative analysis of all the different cryptographic algorithms reveals many facts about cryptographic algorithms execution on cloud network. Some of these algorithms, like homomorphic algorithms, have to pass a long way before their actual implementation on cloud. My further research work is to test more homomorphic schemes which may be used as zero knowledge proof algorithms on cloud network.